# General expressions for the electrical resistivity and thermal conductivity of twinned crystals

V. W. Wittorff[*]

*IRC in Superconductivity and Department of Physics,
University of Cambridge, Madingley Road, Cambridge CB3 0HE, UK*

General expressions are derived for the electrical resistivity $\rho$ and thermal conductivity $\kappa$ of a twinned single crystal. Particular attention is paid to the effect of the structure of the twin domains on these transport coefficients. Edge effects are also considered. The expression for $\kappa$ is used to fit thermal conductivity data for a twinned single crystal of $YBa_2(Zn_{0.008}Cu_{0.992})_3O_{6.98}$. The expression for $\rho$ is used to fit previously published electrical resistivity data for a twinned single crystal of $YBa_2Cu_3O_{6.9}$. It is found that twin boundaries are not a significant source of electron scattering in high-quality single crystals of $YBa_2Cu_3O_{7-\delta}$. We cannot rule out phonon scattering by twin boundaries in these crystals, with up to 12% suppression of the phonon component of $\kappa$. The related problem of determining $\rho$ and $\kappa$ of a crystal with oblique and alternating isotropic regions of different conductivities is also solved.

## I.   INTRODUCTION

Many anisotropic crystalline solids exhibit twinning. For example, the much-studied cuprate superconductor $YBa_2Cu_3O_{7-\delta}$ is orthorhombic, and single crystals of this and other rare-earth-123 materials are often twinned in the *ab*-plane. In the experimental research record there is a large body of work regarding transport properties of these twinned crystals, and it is continuing to grow. However there is presently no published way to reconcile or quantitatively compare the twinned data with the principal components of the transport coefficient tensors (e.g. $\sigma_a$ and $\sigma_b$; or $\kappa_a$ and $\kappa_b$) determined from measurements of untwinned crystals of the same materials. This shortcoming is rectified in the present paper.

Furthermore, the simple arithmetic average of $\rho_a$ and $\rho_b$ has been used several times in the literature[1,2] to attempt (unsuccessfully) to fit twinned electrical resistivity data, whereas use of the exact expression, which is derived in this paper, enables those twinned data of Ref. 1 to be reconciled perfectly with detwinned data of the same crystals (see §IV). Papers from two other groups[3,4] state that "measured quantities in the *a-b* plane are averaged over the *a* and *b* direction." Both of these studies show electrical resistivity data $\rho_{ab}$ for twinned Zn-doped Y-123, and if these twinned data are analysed to extract the corresponding additional planar resistivity per at.%Zn using a simple weighted average for $\rho_{ab}$, the value obtained is only about two-thirds of the true value obtained using the exact expression, which is derived in this paper. Such analyses are discussed in §V.

With regard to the thermal conductivity, Zeini et al.[5] derive an appealing model-independent way to separate the magnetic field-dependent and field-independent parts of the total thermal conductivity. The analysis is on $\kappa_{ab}$ data from a twinned crystal. Unfortunately they make flawed associations between the two separated parts and the physical contributions to the thermal conductivity, as discussed in §IV. Their analysis is flawed because it relies on $\kappa_{ab}$ being comprised of a sum of the physical contributions, which is inconsistent with the expression derived in this paper. In fact the various physical contributions combine in a highly non-linear fashion to give the field-dependent and field-independent parts.

Also it is stated or implied in papers by both Ando's group[6] and Uchida's group[7,8] that measurements on detwinned crystals are necessary in order to extract $\rho_a$, unless anisotropy can be ignored. This latter point is important: attempted use of the simple average for $\rho_{ab}$ has not been successful in the past in extracting the components $\rho_a$ and $\rho_b$, and consequently most authors have concluded that twinned data are of limited use in this respect and measurements of detwinned crystals are required for this purpose. This is not the case. Use of the exact expressions derived in this paper in principle allows twinned data to reveal the $\rho_a$ and $\rho_b$ components unambiguously, provided something is known of the nature of their temperature dependences.

This is not to encourage experimentalists to measure twinned crystals rather than taking the care to detwin them. However the fact is that there is a large body of published data for twinned crystals which is not given the attention it deserves, because of the hitherto lack of exact analytical expressions for its analysis. Even worse is the unnecessary lack of confidence in the twinned data due to authors attempting unsuccessfully to analyse them using the simple average expression for the electrical resistivity. This paper rectifies those misapprehensions.

## II. TWINNED CRYSTAL

We are interested in calculating the electrical resistivity (or thermal conductivity) in a particular direction, of a twinned crystal, when $\rho_a$ and $\rho_b$ (or $\kappa_a$ and $\kappa_b$) are known. The situation is shown in Fig. 1. In practice the crystals are generally long (typical in-plane aspect ratio of 3) and thin. The measurement would be taken along the long axis (parallel to $y$ in Fig. 1) using a traditional four-point technique. Looking at the crystal face, as in Fig. 1, there are alternating regions of crystal orientation – the first where the crystallographic $a$-axis is parallel to the long axis, and the $b$-axis is perpendicular (region type A); the next where the *b*-axis is parallel to the long axis, and the *a*-axis is perpendicular (region type B). Only two such adjacent regions are shown in Fig. 1.

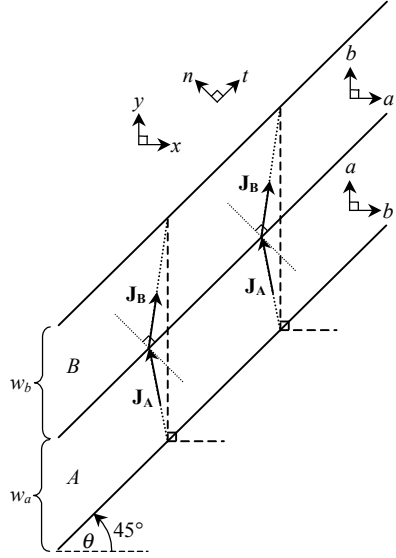

Fig. 1: Pair of adjacent domains in a twinned crystal. Refraction occurs at each boundary. The net global current must be in the *y*-direction.

We can assume without loss of generality (see §VIII) that all of the regions type A are of the same width ($w_a$) and that all of the regions type B are of the same width ($w_b$), but that in general $w_a$ is not necessarily equal to $w_b$. We can also assume without loss of generality that these regions (actually twin domains) stretch across the crystal from one edge of the crystal to another. We ignore edge effects (which are considered in §X) by assuming that $w_a$ and $w_b$ are much smaller than the length and width of the crystal, so that there are many twin domains. Finally, we assume that the interfaces between the regions (the twin boundaries) do not contribute significantly to scattering (this assumption is examined in §IV).

The angle $\theta$ between the normal to the twin boundaries, and the long crystal axis, is shown in Fig. 1 to be 45°. This is always the case, and is a property of twinning.[9] However if we imagine the case where $\theta = 0°$, so that the twin boundaries are normal to the direction of measurement, it is obvious that the current densities $\mathbf{J}_A$ and $\mathbf{J}_B$ will flow parallel to *y* in each domain, thereby being normally incident to each twin boundary. It is intuitively obvious and easy to show that for this case, where $\theta = 0°$, the measured electrical resistivity or thermal conductivity would be that of a series combination of domains, thus:

$$\rho = p\rho_a + (1-p)\rho_b \qquad (1)$$

and
$$\frac{1}{\kappa} = \frac{p}{\kappa_a} + \frac{1-p}{\kappa_b}$$

i.e.
$$\kappa = \frac{\kappa_a \kappa_b}{p\kappa_b + (1-p)\kappa_a} \qquad (2)$$

where $p \equiv w_a/(w_a + w_b)$ is the proportion of crystal area with $a \parallel y$ (that is, of regions type A). Naturally $1 - p \equiv w_b/(w_a + w_b)$ is the proportion of crystal area of regions type B.

However for any case where $\theta \neq 0°$, it is apparent that the electrical or heat current must be deflected on traversing each twin boundary, exactly analogously to the refraction of light. So for charge transport, $\mathbf{J}_A$ will not be parallel to $\mathbf{J}_B$, and the current will zig-zag from one end of the crystal to the other. Let us now analyse this situation, initially for general $\theta$. We will restrict our attention for the time being to an electrical measurement.

At each twin boundary, we have the following entirely general boundary conditions in the steady-state:

$$J_{A,n} = J_{B,n} \qquad (3)$$
$$E_{A,t} = E_{B,t} \qquad (4)$$

Equation 3, that the component of the current density normal to the boundary is continuous across the boundary, is a necessary consequence of conservation of charge. Equation 4, that the component of the electric field tangential to the boundary is continuous across the boundary, is a necessary consequence of a conservative field. (Analogous boundary conditions are applicable for a thermal measurement, in which case $\nabla T$ is the conservative field.) It may be instructive to mention that this second boundary condition is not usually given in the form of Equation 4 in treatises and textbooks considering current flow across an oblique interface between dissimilar media. This is because most analyses are for isotropic media, where it is appropriate to use the less general boundary condition that the product of the conductivity and the tangential current density, is continuous. This is not the case here.

Conservation of charge also requires that the total net current $\mathbf{I} \parallel y$, that is, that $I_x = 0$, since the current contacts would be aligned along the *y*-direction, at opposite ends of the crystal. This constraint is depicted in Fig. 1 by the vertical dashed lines – which serve to demonstrate graphically that when the current vectors $\mathbf{J}_A$ and $\mathbf{J}_B$ are taken to pass through a common point on a twin boundary, extensions in the directions of the two vectors intersect the other boundary of the respective twin domains at vertically-aligned points.

In terms of the current densities, which are different in the two region types (in magnitude *and* direction), this constraint can be most conveniently expressed not by reference to the geometry of the construction in Fig. 1, but instead by projecting the average of the normal- and tangential-components onto the *x*-axis, and equating them thus:

$$J_{B,n} \sin\theta = J_t^{avg} \cos\theta \tag{5}$$

where
$$J_t^{avg} = \frac{w_a J_{A,t} + w_b J_{B,t}}{w_a + w_b} \tag{6}$$

is the value of the tangential component of the current density, averaged over the crystal; and where we could have arbitrarily used $J_{A,n}$ in place of $J_{B,n}$ as the average of $J_n$ in Equation 5, since the two are equal according to Equation 3.

To determine the overall electrical resistivity of the crystal, we divide the average electric field in the direction of net current flow, by the average value of the current density (in this direction), that is:

$$\rho = \frac{E_y^{avg}}{J_y^{avg}} \tag{7}$$

where
$$E_y^{avg} = \frac{w_a E_{A,y} + w_b E_{B,y}}{w_a + w_b} \tag{8}$$

and
$$J_y^{avg} = J_{B,n} \cos\theta + J_t^{avg} \sin\theta \tag{9}$$

We see that for $\theta = 0°$, Equations 7-9 immediately give Equation 1. In general however, we must now introduce Ohm's Law for an anisotropic material:

$$\mathbf{J_A} = \begin{bmatrix} \sigma_b & 0 \\ 0 & \sigma_a \end{bmatrix} \begin{bmatrix} E_{A,x} \\ E_{A,y} \end{bmatrix} = \begin{bmatrix} \sigma_b E_{A,x} \\ \sigma_a E_{A,y} \end{bmatrix} \tag{10}$$

and
$$\mathbf{J_B} = \begin{bmatrix} \sigma_a & 0 \\ 0 & \sigma_b \end{bmatrix} \begin{bmatrix} E_{B,x} \\ E_{B,y} \end{bmatrix} = \begin{bmatrix} \sigma_a E_{B,x} \\ \sigma_b E_{B,y} \end{bmatrix} \tag{11}$$

which gives for the normal and tangential components of the current densities:

$$\left. \begin{array}{l} J_{A,n} = \mathbf{J_A} \bullet \hat{\mathbf{a}}_\mathbf{n} = -\sigma_b E_{A,x} \sin\theta + \sigma_a E_{A,y} \cos\theta \\ J_{B,n} = \mathbf{J_B} \bullet \hat{\mathbf{a}}_\mathbf{n} = -\sigma_a E_{B,x} \sin\theta + \sigma_b E_{B,y} \cos\theta \\ J_{A,t} = \mathbf{J_A} \bullet \hat{\mathbf{a}}_\mathbf{t} = \sigma_b E_{A,x} \cos\theta + \sigma_a E_{A,y} \sin\theta \\ J_{B,t} = \mathbf{J_B} \bullet \hat{\mathbf{a}}_\mathbf{t} = \sigma_a E_{B,x} \cos\theta + \sigma_b E_{B,y} \sin\theta \end{array} \right\} \tag{12}$$

Substituting Equations 12 into 3 gives:

$$-\sigma_b E_{A,x} \sin\theta + \sigma_a E_{A,y} \cos\theta = -\sigma_a E_{B,x} \sin\theta + \sigma_b E_{B,y} \cos\theta \tag{13}$$

Equation 4 gives:

$$E_{A,x} \cos\theta + E_{A,y} \sin\theta = E_{B,x} \cos\theta + E_{B,y} \sin\theta \tag{14}$$

and substituting Equations 6 and 12 into 5 gives:

$$-\sigma_a E_{B,x} \sin^2\theta + \sigma_b E_{B,y} \sin\theta \cos\theta$$
$$= \frac{w_a \sigma_b E_{A,x} \cos\theta + w_a \sigma_a E_{A,y} \sin\theta + w_b \sigma_a E_{B,x} \cos\theta + w_b \sigma_b E_{B,y} \sin\theta}{w_a + w_b} \tag{15}$$

Now making the substitution $\theta = 45°$, Equations 13-15 can be written as the system of linear equations:

$$\begin{bmatrix} -\sigma_b & \sigma_a & \sigma_a & -\sigma_b \\ 1 & 1 & -1 & -1 \\ w_a \sigma_b & w_a \sigma_a & w_a \sigma_a + 2 w_b \sigma_a & -w_a \sigma_b \end{bmatrix} \begin{bmatrix} E_{A,x} \\ E_{A,y} \\ E_{B,x} \\ E_{B,y} \end{bmatrix} = \begin{bmatrix} 0 \\ 0 \\ 0 \end{bmatrix} \tag{16}$$

Solving in terms of $E_{B,y}$, we have:

$$E_{B,x} = E_{B,y} \frac{\sigma_b (\sigma_b - \sigma_a)}{\sigma_a \left[ \frac{w_b}{w_a}(\sigma_a + \sigma_b) + 2\sigma_b \right]}, \tag{17}$$

$$E_{A,y} = E_{B,y} \frac{\sigma_b \left(\sigma_b + \sigma_a + 2\frac{w_b}{w_a}\sigma_a\right)}{\sigma_a \left[ \frac{w_b}{w_a}(\sigma_a + \sigma_b) + 2\sigma_b \right]}, \tag{18}$$

and

$$E_{A,x} = E_{B,y} \frac{\frac{w_b}{w_a}(\sigma_a - \sigma_b)}{\frac{w_b}{w_a}(\sigma_a + \sigma_b) + 2\sigma_b} \tag{19}$$

Substituting Equation 18 into 8 gives:

$$E_y^{avg} = \frac{E_{B,y}}{w_a + w_b} \frac{(\sigma_a + \sigma_b)\left(w_a \sigma_b + \frac{1}{w_a} w_b^2 \sigma_a\right) + 4 w_b \sigma_a \sigma_b}{\sigma_a \left[ \frac{w_b}{w_a}(\sigma_a + \sigma_b) + 2\sigma_b \right]} \tag{20}$$

and substituting Equation 17 into 12 gives for $J_{B,n}$ with $\theta = 45°$:

$$J_{B,n} = \frac{E_{B,y}}{\sqrt{2}} \frac{\sigma_b (\sigma_a + \sigma_b)\left(1 + \frac{w_b}{w_a}\right)}{\frac{w_b}{w_a}(\sigma_a + \sigma_b) + 2\sigma_b} \tag{21}$$

For $\theta = 45°$, Equation 5 requires that $J_t^{avg} = J_{B,n}$, which substituted into Equation 9 gives:

$$J_y^{avg} = \sqrt{2} J_{B,n} \tag{22}$$

Substituting Equations 20-22 into 7 gives the electrical resistivity, which written in terms of $p$ is:

$$\rho = \frac{[p\sigma_b + (1-p)\sigma_a]^2 + \sigma_a \sigma_b}{\sigma_a \sigma_b (\sigma_a + \sigma_b)} \tag{23}$$

Written in terms of $\rho_a$ and $\rho_b$, this is:

$$\rho = \frac{[p\rho_a + (1-p)\rho_b]^2 + \rho_a \rho_b}{\rho_a + \rho_b} \tag{24}$$

Similarly, the total thermal conductivity is:

$$\kappa = \frac{\kappa_a \kappa_b (\kappa_a + \kappa_b)}{[p\kappa_b + (1-p)\kappa_a]^2 + \kappa_a \kappa_b} \tag{25}$$

It is interesting that for an "equally" twinned crystal, i.e. with $p = 0.5$, Equation 24 reduces to:

$$\rho = \frac{\rho_a + \rho_b}{4} + \frac{\rho_a \rho_b}{\rho_a + \rho_b} = \frac{\rho_{series} + \rho_{parallel}}{2} \tag{26}$$

where $\rho_{series}$ is the resistivity of a series combination of parallelepipeds of equal geometry having resistivities of $\rho_a$ and $\rho_b$ respectively, and $\rho_{parallel}$ is the resistivity of a parallel combination of the same two parallelepipeds.

### III.    JOULE HEATING AND ITS ANALOGUE

During an electrical measurement, the power dissipated per unit volume (the power density) due to Joule heating is given by $\mathbf{J} \bullet \mathbf{E}$. We may determine the local power density in each type of region as follows:

In regions type A:    $\mathbf{J_A} \bullet \mathbf{E_A} = \sigma_b E_{A,x}^2 + \sigma_a E_{A,y}^2$    (27)

In regions type B:    $\mathbf{J_B} \bullet \mathbf{E_B} = \sigma_a E_{B,x}^2 + \sigma_b E_{B,y}^2$    (28)

The overall measured power density, as determined from the measured voltage and current and the crystal dimensions, is given by $J_y^{avg} E_y^{avg}$. By substituting Equations 20-22 into this expression; and substituting Equations 17-19 into expressions 27 and 28; we find:

$$\mathbf{J_A} \cdot \mathbf{E_A} = \mathbf{J_B} \cdot \mathbf{E_B} = J_y^{avg} E_y^{avg} \tag{29}$$

So the power dissipated per unit volume due to Joule heating is the same throughout the crystal. Furthermore we have shown that it is equal to the value obtained using the measured voltage and current, as must be the case.

An analogous result is found for a thermal measurement, for the product of the heat current $\mathbf{J}^q$ (units of $Wcm^{-2}$) and the temperature gradient $\nabla T$ (units of $Kcm^{-1}$):

$$\mathbf{J_A^q} \cdot (\nabla T)_A = \mathbf{J_B^q} \cdot (\nabla T)_B = J_y^{q,avg} (\nabla_y T)^{avg} \tag{30}$$

## IV. USE OF THE EXPRESSIONS WITH MEASURED DATA

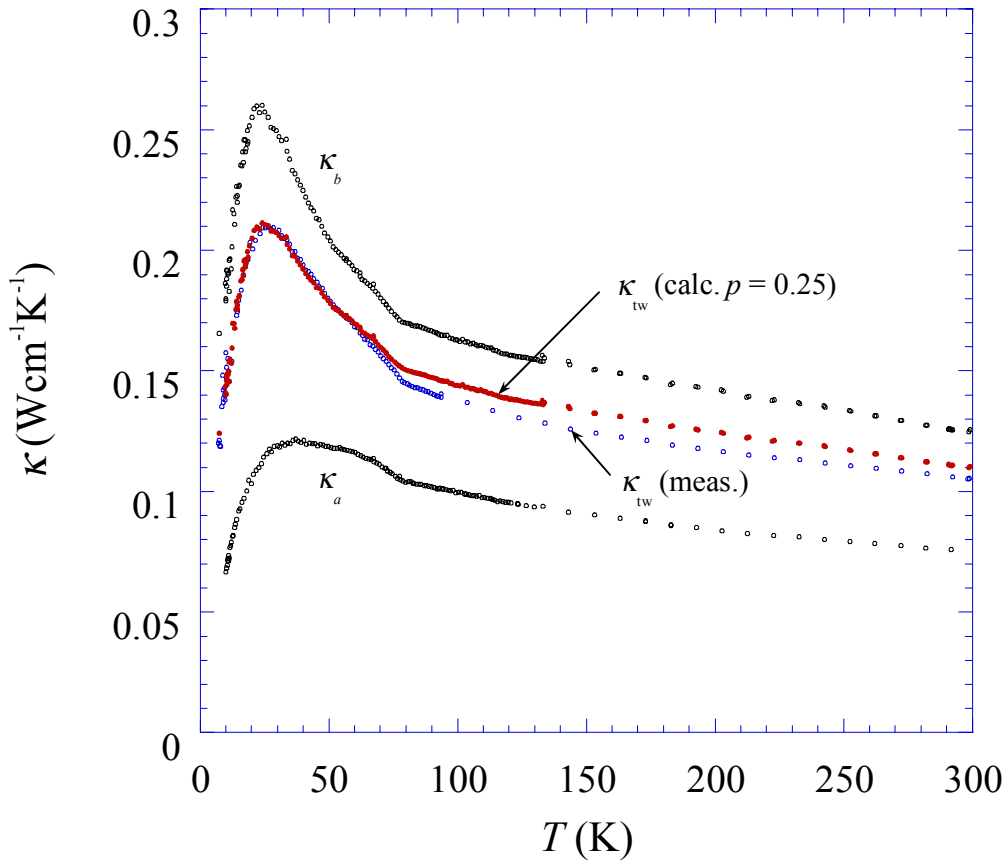

Fig. 2: (Color online) Fitting thermal conductivity data $\kappa_{tw}$ of twinned crystal, with combination of data of detwinned crystals, for 0.8% Zn-doped Y-123 single crystals.

We have performed measurements in the manner described above, of the thermal conductivity of detwinned and twinned single crystals of $YBa_2(Zn_xCu_{1-x})_3O_{6.98}$. Data from these measurements and associated analysis are reported in detail elsewhere.[10,11,12] We show in Fig. 2 data for crystals with Zn-doping of $x = 0.008$. Measurement of detwinned crystals has provided $\kappa_a$ and $\kappa_b$, as shown. Data from the measurement of a twinned crystal are also shown, and are adequately fit using Equation 25 with $p = 0.25$. This value for $p$ is consistent with the proportion of the twinned crystal's area comprised of regions type A, determined empirically using a polarising microscope.

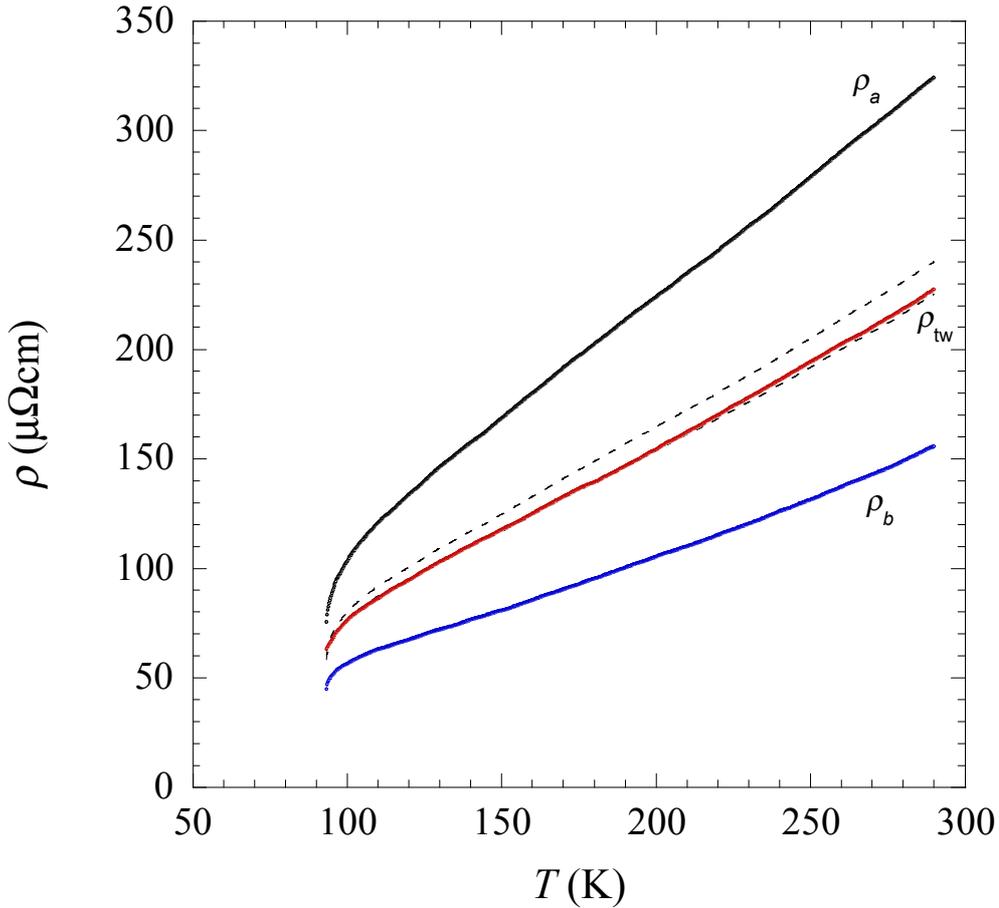

Fig. 3: (Color online) Fitting electrical resistivity data $\rho_{tw}$ of twinned crystal, with combination of data of the same crystal after detwinning, for a Y-123 single crystal. Data are from Ref. 1, and appear as thick solid traces. Two fits to $\rho_{tw}$ are shown as thin dashed traces. The upper fit is $(\rho_a + \rho_b)/2$, as given in Ref. 1. The lower fit overlaps the $\rho_{tw}$ data and is Equation 24 with $p = 0.5$.

In Fig. 3 are shown data from Gagnon et al.[1], of $\rho_a$ and $\rho_b$ for a single crystal sample (Crystal No. 1) of $YBa_2Cu_3O_{6.9}$. The data were obtained using the Montgomery method[13]. Also shown is the electrical resistivity $\rho_{tw}$ of the same crystal prior to detwinning. Gagnon et al. state that after oxygenation and before detwinning, the crystal was microtwinned almost uniformly over its entire surface (so $p = 0.5$). Assuming that twin boundaries do not increase the planar resistivity, they state that the electrical resistivity of the twinned crystal $\rho_{tw}$ would be the simple average of $\rho_a$ and $\rho_b$ (that is, Equation 1 with $p = 0.5$). This is not the case according to the analysis presented in §II of this paper. The attempted fit using this expression is shown in Fig. 3; for all temperatures the calculated values are too large by 6-7%. Gagnon et al. attribute this discrepancy to geometric errors, despite their convincing description of the lengths to which they have gone to ensure the accuracy of the magnitudes of all of their data. However the present analysis suggests that their data are correct, and it is their attempted fit that is flawed.

We also show in Fig. 3 the fit to the twinned data using our general expression 24 with $p = 0.5$ (equivalent to the mnemonic Equation 26). For all temperatures the calculated values are within 1% of the data – for most of the temperature range the fit is completely masked by the data.

The value of $(\rho_a + \rho_b)/2$ is always higher than that of Equation 24 with $p = 0.5$. This discrepancy is larger the higher the anisotropy. For twinned $YBa_2Cu_3O_x$, this means that the discrepancy is larger for higher oxygenation. We have shown above using the data of Gagnon et al.[1] that for $x = 6.9$, the discrepancy is 6-7%. Using the detwinned data of Segawa et al.[14], we predict that for $x = 6.95$, the discrepancy would be 10-11%, and for $x = 7.00$, the discrepancy would be 12-13%.

The success of the fit to the twinned electrical resistivity data in Fig. 3 using Equation 24 bears out conclusively our earlier assumption that twin boundaries are not a significant scattering mechanism for an electrical measurement. This means that twin boundaries do not significantly scatter electrons in these high-quality single crystals.

While the fit to the twinned thermal conductivity data in Fig. 2 using Equation 25 is adequate, and convincingly fits the shape of the superconducting enhancement peak, it could not be done while simultaneously fitting the normal state data perfectly. This might be because the data for $\kappa_a$ were taken on a different crystal detwinned from the same batch, which might have had a slightly lower magnitude for $\kappa_a$ than that of the crystal from which both the twinned and *b*-axis data were obtained. However since the calculated values exceed the measured twinned data in the normal state, we cannot rule out the possibility that the twin boundaries act as a source of scattering in a thermal measurement. In this case the suppression is about 7% throughout the normal state. If we assume that – as for the electrical measurement – electrons are not scattered by the twin boundaries, then we can attribute the entire suppression to the phonon component $\kappa^{ph}$ of the thermal conductivity. In this case the suppression due to scattering of phonons by twin boundaries corresponds to about 12% of $\kappa^{ph}$ throughout the normal state.[10] We may consider this an upper bound.

Zeini et al.[5] describe a model-independent way to separate the magnetic field-dependent and field-independent components of the total thermal conductivity, whence $\kappa = \kappa^{B\text{-dep}} + \kappa^{B\text{-indep}}$, with $\kappa^{B\text{-dep}}(B \to \infty) = 0$. The analysis is on $\kappa_{ab}$ data from a twinned crystal. They equate $\kappa^{B\text{-dep}}$ (which they call $\kappa_{xx}^{el}$) with solely the electronic plane contribution, and equate $\kappa^{B\text{-indep}}$ (which they call $\kappa_{xx}^{rest}$) with the simple sum of the chain and phonon contributions. In fact neither association is more than a rough approximation, and this is clear when the expression (Equation 25) is examined for the total thermal conductivity of a twinned crystal. The expression is a highly non-linear combination of (i) $\kappa_a = \kappa_{pl}^{el} + \kappa_a^{ph}$, and (ii) $\kappa_b = \kappa_{pl}^{el} + \kappa_{ch}^{el} + \kappa_b^{ph}$, and the resulting highly non-linear expression for $\kappa^{B\text{-dep}}$ involves all five contributions, not just $\kappa_{pl}^{el}$ (which for typical values might be 10% smaller than $\kappa^{B\text{-dep}}$, but this discrepancy is temperature-dependent). Unfortunately Zeini *et al.* do claim to be able to extract the field-dependent $\kappa_{pl}^{el}$ in this way, by assuming that

$\kappa_{ab} = \kappa_{pl}^{el} + \langle \kappa_{ch}^{el} \rangle + \langle \kappa^{ph} \rangle$, where the $\langle . \rangle$ denote some kind of average appropriate for a twinned crystal. This is a fundamental flaw in their analysis, since any sum of this kind is inconsistent with the expression for $\kappa_{ab}$ derived in this paper. On the other hand, their method of analysis would be able to separate the different physical contributions to the thermal conductivity for an *untwinned* crystal, as long as the particular empirical observation on which their entire analysis hinges also applies to data from untwinned crystals.

Finally, we note that the same expression for $\rho_{tw}$ that is used by Gagnon *et al.*, is used by Terasaki *et al.*[2] to calculate the expected resistivity of twinned thin films.

## V. "MATTHIESSEN'S RULE" FOR $\rho_{ab}$

It is sometimes claimed[3,4,15] that the electrical resistivity of Zn-doped twinned Y-123 crystals obeys Matthiessen's Rule, in that there appears to be a constant offset between the curves (additional residual resistivity) proportional to the level of Zn. The physical justification for this is indirect. We might expect Matthiessen's Rule to be obeyed for $\rho_{pl}$ of Zn-doped Y-123, since Zn is thought to substitute for the planar Cu(2) atoms. However the effect of Zn on $\rho_{ch}$ might be expected to be far weaker or even zero, in the absence of Zn in the chains. Now $\rho_b = \left(1/\rho_{pl} + 1/\rho_{ch}\right)^{-1}$, and consequently the increase with Zn doping of $\rho_b$ is proportionately far weaker than for $\rho_a$ for even low Zn levels, and for high Zn levels the $\rho_b$ values quickly asymptote towards $\rho_{ch}$.

If we are interested in $\rho_{ab}$ of an equally twinned crystal ($p = 0.5$), we might be motivated by the statements in Refs 3 and 4 to represent it by the simple average $(\rho_a + \rho_b)/2$, which we have pointed out in §IV above always gives values that are too large. From the author's own measurements[10] of detwinned YBa$_2$(Zn$_{0.008}$Cu$_{0.992}$)$_3$O$_{6.98}$, and measurements of detwinned YBa$_2$Cu$_3$O$_{6.98}$ from Ref. 16, $\rho_a$ obeys Matthiessen's rule precisely, increasing by $33.4\mu\Omega\text{cm/at.\%Zn}$. Since $\rho_b$ increases significantly more weakly as the level of Zn is increased – when it is assumed that $\rho_{ch}$ is unaffected by Zn – we would expect that the $\rho_{tw}$ values calculated according to this expression (the simple average) to vary with Zn with slightly more than half the sensitivity as that of $\rho_a$. This is precisely what we find, as shown by the upper set of curves in Fig. 4, which give a sensitivity to Zn of about $18\,\mu\Omega\text{cm/at.\%Zn}$.

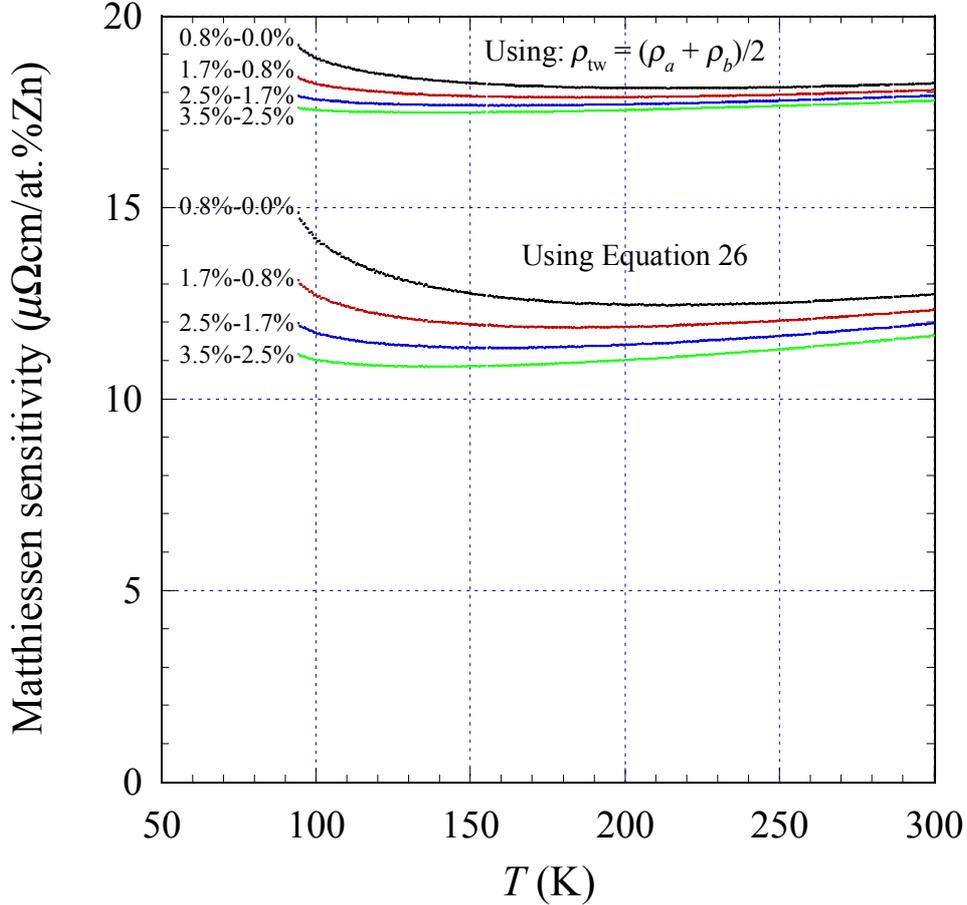

Fig. 4: (Color online) Matthiessen sensitivities for twinned YBa$_2$(Cu$_{1-x}$Zn$_x$)$_3$O$_{6.98}$, using $\rho_{tw} = (\rho_a + \rho_b)/2$ and using Equation 26, with $p = 0.5$. (See text.)

Fig. 4 requires explanation. Values for $\rho_{tw}$ are not shown, but rather their sensitivity to Zn. That is the increase in $\rho_{tw}$ from one Zn-doping level to the next, relative to the increase in the Zn content, that is:

$$\text{Matthiessen sensitivity} = \frac{\rho_{tw}(T; x_2(\text{Zn})) - \rho_{tw}(T; x_1(\text{Zn}))}{100 \times [x_2(\text{Zn}) - x_1(\text{Zn})]}\,\mu\Omega\text{cm/at.\%Zn} \quad (31)$$

The upper set of curves is for $\rho_{tw}$ represented by $(\rho_a + \rho_b)/2$. The facts that they are essentially temperature independent, and vary only slightly on the levels of Zn, show that the curves generated by $(\rho_a + \rho_b)/2$ could be described as "obeying Matthiessen's Rule".

The lower set of curves is for $\rho_{tw}$ represented by the true expression derived in this paper and given by Equation 26. They have more temperature dependence than the upper set of curves, but can still be described as approximately temperature independent. They vary more strongly with the levels of Zn than the upper set of curves, but can still be described as approximately independent of the absolute levels of Zn. So the twinned data could be described as "obeying Matthiessen's Rule" to a reasonable approximation. However, the striking feature of Fig. 4 is the difference in the overall magnitude of the two sets of curves. The upper set gives a sensitivity of about $18\mu\Omega$cm/at.%Zn, while the lower set gives a sensitivity of only about $12\mu\Omega$cm/at.%Zn. In other words, using the expression $(\rho_a + \rho_b)/2$ will give a Matthiessen sensitivity that is 50% too high for $YBa_2Cu_3O_{6.98}$, while the magnitude of the electrical resistivity of such a crystal calculated in this way is (only) 12% too high (see §IV). Conversely, if data for twinned Zn-doped $YBa_2Cu_3O_{6.98}$ crystals were analysed to extract the corresponding *planar* Matthiessen sensitivity using a simple average for $\rho_{ab}$, the value obtained would only be two-thirds of the true value.

This is remarkable, and the degree of discrepancy of the Matthiessen sensitivity is perhaps unexpected. It shows that the *derivatives* (with respect to anisotropy $\rho_b/\rho_a$) of the two expressions for $\rho_{tw}$ are very different near $p = 0.5$, since this derivative is what determines the Matthiessen sensitivity.

## VI. THE CASE OF ISOTROPIC DOMAINS

A related problem involves alternating regions similar to those shown in Fig. 1, but where regions type A are each isotropic with electrical resistivity $\rho_a$ (and thermal conductivity $\kappa_a$), and where regions type B are each isotropic with $\rho_b$ (and $\kappa_b$). The analysis proceeds along the same lines as for the twinned crystal with anisotropic domains, but Equations 10 and 11 are replaced with:

$$\mathbf{J_A} = \begin{bmatrix} \sigma_a & 0 \\ 0 & \sigma_a \end{bmatrix} \begin{bmatrix} E_{A,x} \\ E_{A,y} \end{bmatrix} = \sigma_a \mathbf{E_A} \quad (32)$$

and

$$\mathbf{J_B} = \begin{bmatrix} \sigma_b & 0 \\ 0 & \sigma_b \end{bmatrix} \begin{bmatrix} E_{B,x} \\ E_{B,y} \end{bmatrix} = \sigma_b \mathbf{E_B} \quad (33)$$

which affects the subsequent working. The final equation for the electrical resistivity is:

$$\rho = \frac{p(1-p)(\rho_a - \rho_b)^2 + 2\rho_a \rho_b}{2[p\rho_b + (1-p)\rho_a]} \quad (34)$$

and for the total thermal conductivity is:

$$\kappa = \frac{2\kappa_a\kappa_b[p\kappa_a + (1-p)\kappa_b]}{p(1-p)(\kappa_b - \kappa_a)^2 + 2\kappa_a\kappa_b} \quad (35)$$

These expressions in general are unequal to Equations 24 and 25 respectively. Of course for the trivial cases where $p = 1$ or $0$, corresponding to an untwinned crystal (or absence of dissimilar domains), both pairs of equations reduce to the same expressions:

$$\rho = \begin{cases} \rho_a, & p = 1 \\ \rho_b, & p = 0 \end{cases} \quad (36)$$

$$\kappa = \begin{cases} \kappa_a, & p = 1 \\ \kappa_b, & p = 0 \end{cases} \quad (37)$$

It is more interesting that for $\theta = 45°$ and $p = 0.5$, both pairs of equations also reduce to the same expressions, but for no other non-trivial value of $p$. In this special case of an "equally" twinned (or striped) crystal, Equation 26 gives the electrical resistivity whether the domains are anisotropic or not.

## VII. THE CASE OF AVERAGE CURRENT DENSITY OBLIQUE TO CRYSTALLOGRAPHIC AXES

In the preceding analysis, the current contacts were aligned in the $y$-direction, which was parallel to the long axis of the crystal and a principal crystallographic axis direction. This is the usual way in which a measurement is performed.

Let us instead consider the situation where the current contacts are not aligned with the principal crystallographic axes. For example, a parallelepiped could be cut obliquely out of a crystal, so that when the current contacts are aligned with the long axis of this parallelepiped, they are not aligned with the crystallographic axes. Let us denote this angle of misalignment as $\phi$, and let us rotate by $\phi$ the $x$-$y$ coordinate system shown in Fig. 1 so that the $y$-axis is again parallel to the long axis of this crystal. Now, Equations 3-9 are all still applicable in this situation, except $\theta$ must be replaced with $\theta \pm \phi$, as the case may be.

For the case of isotropic regions, Equations 32 and 33 for Ohm's Law are still applicable, and the analysis proceeds as before, but with $\theta \pm \phi$ instead of $\theta$ as the angle between the normal $n$ to the region boundaries, and the $y$-axis.

However for the case of anisotropic regions (a twinned crystal), Equations 10 and 11 no longer describe Ohm's Law. If we denote by the vectors $\mathbf{J}'$ and $\mathbf{E}'$ the current density and electric field (respectively) in the new re-aligned $x$-$y$ coordinate system, then in terms of $\mathbf{J}$ and $\mathbf{E}$ in the old $x$-$y$ coordinate system they are given by:

$$\mathbf{J}' = \mathbf{R}(\phi)\mathbf{J} \tag{38}$$

and
$$\mathbf{E}' = \mathbf{R}(\phi)\mathbf{E} \tag{39}$$

where
$$\mathbf{R}(\phi) = \begin{bmatrix} \cos\phi & \sin\phi \\ -\sin\phi & \cos\phi \end{bmatrix} \tag{40}$$

is the rotation operator. In the new *x-y* coordinate system, Ohm's Law is given by $\mathbf{J}' = \bar{\sigma}'\mathbf{E}'$, where the relationship between the conductivity tensor $\bar{\sigma}'$ in this rotated coordinate system, and $\bar{\sigma}$ in the old (crystallographically-aligned) *x-y* coordinate system, is given by:

$$\bar{\sigma}' = \mathbf{R}(\phi)\,\bar{\sigma}\,\mathbf{R}(-\phi) \tag{41}$$

These transformed conductivity tensors will no longer be diagonal. With Equations 10 and 11 rewritten using these transformed conductivity tensors, and with $\theta$ replaced with $\theta \pm \phi$ as the case may be, the analysis proceeds as before.

## VIII.  THE CASE OF AN "UNEVENLY" TWINNED CRYSTAL

An early assumption made in our analysis is that all regions type A are of width $w_a$, and that all regions type B are of width $w_b$. Consequently the pair of adjacent domains shown in Fig. 1 is perfectly representative of all others, and the crystal twin domain structure is obtained by simply replicating this pair of domains as many times as required.

We may describe such a crystal as being "evenly" twinned, which does *not* mean that $p = 0.5$ necessarily[17], but rather that the local proportion $p$ of the crystal area comprised of regions type A is the same in any chosen sub-area of the crystal's surface.

If we relax this assumption, we have the more general case of an "unevenly" twinned crystal. In this case, the local value of $p$ varies from one sub-area of the crystal surface to another. A global (average) value of $p$ can still be defined, however. Is this global average value of $p$ sufficient to determine $\rho$ (or $\kappa$), given $\rho_a$ and $\rho_b$ (or $\kappa_a$ and $\kappa_b$)?

The simple answer is yes. This can be easily proved once it is realised that as a result of the twin boundaries being parallel, the current density $\mathbf{J}$ and electric field $\mathbf{E}$ must recover their original magnitudes and directions after traversing any two interfaces (i.e. A→B→A, or B→A→B). Hence $\mathbf{J}$ and $\mathbf{E}$ are the same in all regions type A, and are the same in all regions type B, irrespective of the widths of *any* of the regions, and therefore irrespective of whether the ratios of the widths of adjacent regions varies from one part of the crystal to another.

Now examining Equations 6-9, we see that $\rho$ depends only on the average $y$-components of $\mathbf{J}$ and $\mathbf{E}$. Since we have said that $\mathbf{J}$ and $\mathbf{E}$ in any particular region do not depend on the position in the crystal of the region with respect to other regions, we could arbitrarily re-order the regions without altering $\rho$. Imagine a re-ordering so that all of the regions type A are adjacent, with then a single interface to an aggregation of all of the regions type B. This would result in the situation first depicted in Fig. 1, but with the single region widths $w_a$ and $w_b$ replaced by the *sum* of widths of all regions of each respective type. The distribution of these widths is therefore irrelevant, and it is only the ratio of the sums which affects the calculation. Obviously this ratio is $p/(1-p)$, where here $p$ is the global average.

Thus the previous assumption of an "evenly" twinned crystal was made without loss of generality, and in fact all of the previous analysis and results are equally applicable in the general case of an "unevenly" twinned crystal.

It is necessary though to reconsider the nature of the dashed lines in Fig. 1 for the case of pairs of single (not aggregated) regions in an "unevenly" twinned crystal. In particular, the dashed lines will only be exactly vertical for pairs of adjacent domains that have widths in proportion to the ratio of the respective aggregates, that is, for $w_a/w_b = p/(1-p)$, where here $p$ is the global average. For other (not perfectly representative) pairs of adjacent domains, the directions of $\mathbf{J_A}$ and $\mathbf{J_B}$ are no different, as explained above; but the dashed lines of the kind shown in Fig. 1 will not be exactly vertical, indicating a net transverse drift of current in the pair. However the net transverse drift for pairs of adjacent domains where $w_a/w_b > p/(1-p)$ will be exactly compensated by the net transverse drift in the opposite direction for pairs of adjacent domains where $w_a/w_b < p/(1-p)$. This is guaranteed by the boundary condition that $I_x = 0$ (conservation of charge).

## IX.   THE CASE OF EXTREME INTRA-DOMAIN ANISOTROPY

It has already been mentioned in §IV that for the case $p = 0.5$, the value of the simple weighted average of $\rho_a$ and $\rho_b$ (given by Equation 1) is always higher than that of the true expression for $\rho_{tw}$ given by Equation 24, and that this discrepancy is larger the higher the anisotropy. It is therefore interesting to consider the limit of extreme anisotropy within each domain of the conductivity, for general $p$.

Equation 24 reduces to the following expressions in the two limits of extreme anisotropy:

$$\rho = \begin{cases} p^2 \rho_a, & \rho_b \ll \rho_a \\ (1-p)^2 \rho_b, & \rho_a \ll \rho_b \end{cases} \quad (42)$$

The corresponding expressions for the simple weighted average given by Equation 1 contain exponents of 1 on the geometric factors, instead of 2. For the case of $p = 0.5$, for which the absolute discrepancy between the two pairs of expressions is greatest, the simple average is therefore a factor of 2 too large.

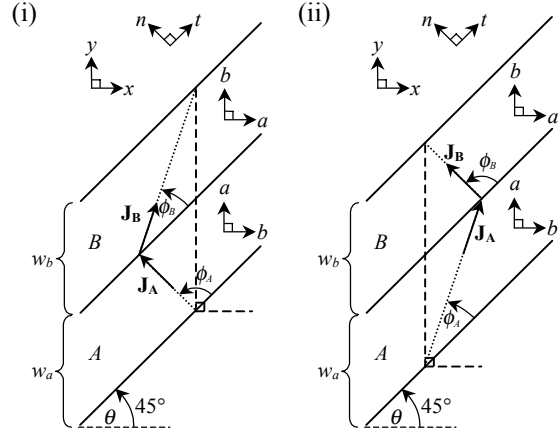

Fig. 5: Relationship between directions of the current densities in a representative pair of adjacent domains (where $w_a/w_b = p/(1-p)$), for (i) $\rho_b \ll \rho_a$, and (ii) $\rho_a \ll \rho_b$. See text for expressions for $\phi_A$ and $\phi_B$. Note that for the situation depicted, the angles have sizes as drawn.

The current densities $\mathbf{J}_A$ and $\mathbf{J}_B$ are parallel to the $y$-axis for isotropic domains, and the deviation of $\mathbf{J}_A$ and $\mathbf{J}_B$ from the $y$-direction is larger the higher the anisotropy. Fig. 5 shows the two limits of extreme anisotropy, where $\mathbf{J}_A$ and $\mathbf{J}_B$ exhibit maximum deviation from the $y$-direction. The angles (denoted $\phi_A$ and $\phi_B$ respectively) at which $\mathbf{J}_A$ and $\mathbf{J}_B$ flow with respect to the twin boundaries can be calculated using Equations 17-19 and 12, where $\tan\phi_A = J_{A,n}/J_{A,t}$, and $\tan\phi_B = J_{B,n}/J_{B,t}$. We give the expressions for general $p$, and the values for $p = 0.5$, in Table 1. Note that since our angular reference is the direction tangential to the twin boundaries (the $t$-direction), an angle of 45° would correspond to the $y$-direction.

Table 1: General expressions for $\phi_A$ and $\phi_B$, and values for $p = 0.5$, in the limits of extreme anisotropy.

| Condition | $\phi_A$ for general $p$ | $\phi_B$ for general $p$ | $\phi_A$ for $p = 0.5$ | $\phi_B$ for $p = 0.5$ |
|---|---|---|---|---|
| $\rho_b \ll \rho_a$ | $\cot^{-1}(2p-1)$ | $\cot^{-1}(2p+1)$ | 90° | 26.57° |
| $\rho_a \ll \rho_b$ | $\cot^{-1}(3-2p)$ | $\cot^{-1}(1-2p)$ | 26.57° | 90° |

The expressions and values given in Table 1 only depend on the global average value of $p$, and the angles $\phi_A$ and $\phi_B$ will be the same in all domains even for an "unevenly" twinned crystal, as explained in §VIII above. However as explained at the end of §VIII, the dashed lines of the kind shown in Fig. 5 will only be exactly vertical for perfectly representative pairs of adjacent domains, that is for which $w_a/w_b = p/(1-p)$.

## X. EDGE EFFECTS

In the preceding analyses, edge effects were ignored. To include edge effects would require an additional boundary condition: that along the edges running in the $y$-direction (parallel to the direction of average current flow), the local current density must be parallel to the edge. To achieve this for an electrical measurement, static excess charge must have accumulated near these edges, to alter the local electric field. The distribution of excess charge in a region type A will be equal and opposite to that in a region type B of the same width.

The influence of these excess charges diminishes as one moves away from the edges. For a twinned crystal of typical dimensions with many domains, the distortion of **J** and **E** from the values obtained earlier will be negligible across almost all of the crystal width. In these circumstances, a thermal conductivity measurement in which thermocouple contacts are made approximately equidistant from each of the long edges of the crystal, will obviously not be affected by edge effects. Even in the type of electrical resistivity measurement where the voltage contacts are made to the edge of the crystal, edge effects will typically play no part, since $E_y^{avg}$ along the edge is unaffected by the accumulated charge (the net excess charge is zero), and $J_y^{avg}$ is a bulk average.

However the quantity of excess charge which accumulates in each region is proportional to the region's width $w_a$ or $w_b$. The influence of this charge therefore extends further into the crystal interior for larger $w_a$ or $w_b$. In the extreme case where $w_a$ or $w_b$ is comparable to, or greater than, the width of the crystal itself, edge effects might dominate across the entire crystal. In this case it could be that $\mathbf{J} \parallel y$ at all points in the crystal, and so also $\mathbf{E} \parallel y$ throughout. The domains would then act as if they were in a simple series combination, and one might expect Equations 1 and 2 to apply.

## XI. ACKNOWLEDGEMENTS

Thanks to R. M. Howard, J. R. Cooper and J. W. Loram for helpful discussions.